# Proactive Quality Guidance for Model Evolution in Model Libraries

Andreas Ganser[1], Horst Lichter[1], Alexander Roth[2], and Berhard Rumpe[2]

[1] RWTH Aachen University, Ahornstr. 55, 52074 Aachen, Germany
{ganser, lichter}@swc.rwth-aachen.de,
home page: http://www.swc.rwth-aachen.de

[2] RWTH Aachen University, Ahornstr. 55, 52074 Aachen, Germany
{roth, rumpe}@se.rwth-aachen.de,
home page: http://www.se.rwth-aachen.de

**Abstract.** Model evolution in model libraries differs from general model evolution. It limits the scope to the manageable and allows to develop clear concepts, approaches, solutions, and methodologies. Looking at model quality in evolving model libraries, we focus on quality concerns related to reusability.

In this paper, we put forward our proactive quality guidance approach for model evolution in model libraries. It uses an editing-time assessment linked to a lightweight quality model, corresponding metrics, and simplified reviews. All of which help to guide model evolution by means of quality gates fostering model reusability.

## 1 The Need for Proactive Quality Guidance

Modeling is one of the traditional disciplines in computer sciences and other sciences. Therefore, computer scientists have been creating models for decades and have seen models incarnate in a lots of different forms. Interestingly enough that the general modeling theory was not developed by a computer scientist but by Herbert Stachowiak in the seventies [Stachowiak, 1973]. His work found impact in a lot of domains in computer sciences, e.g. databases, resulting in research like generic model management [Melnik, 2004]. This transforms Stachowiak's rather abstract theories to an applicable approach offering concepts and algorithms, e.g., `diff`, `merge`, `similarity`, and `match` operations. As a result, pure operations on models did not seem to be challenging any more and a broader perspective was investigated.

A similar development took place in object oriented modeling which brought up UML as a suitable modeling language. Today, the success of UML is often accredited for two reasons. First, UML is believed to be an effective language, because "for larger and distributed projects, UML modeling is believed to contribute to shared understanding of the system and more effective communication" [Chaudron et al., 2012]. Second, UML is considered as the de facto standard in modeling. Due to that, a lot of tools were developed around UML including code generators. They bolster approaches like rapid prototyping or model driven development (MDD) and allow modelers to deal with complexity on an appropriate level of abstraction.

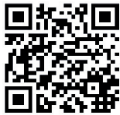

Consequently, UML models are widely used and can be regarded as project assets that should be reused. Moreover, it is believed that model reuse could decrease development time while increasing software quality [Mens et al., 1999, Lange and Chaudron, 2005], because best practices and experience would be leveraged. But the question is how to store models in a way that their quality is maintained or even improved over time. Certainly, model reuse requires an infrastructure enabling to persist models in a library or knowledge base. Furthermore, it needs a means to control model evolution and quality in the long run. Unfortunately, quality is a matter of subjectivity, often relative to requirements and sometimes hard to measure [Moody, 2005]. Moreover, all-at-once quality assessments result in endless quality reports that are hard to work through. One way out is edit-time quality assessment and guidance for assuring a certain level of quality in model libraries, we call proactive quality guidance. To the best of our knowledge we could not find such an approach for model libraries.

Hence, we looked into recent research (section 2) and found that model evolution is often considered as a goal. That would be self-defeating in model libraries. So, we adopted the meaning of model evolution to fit model libraries and developed an approach (section 3) that explains how models should evolve in model libraries. This enables us to discuss model evolution in model libraries on a more formal and qualitative level. In detail, we introduce our understanding of model quality and quality gates. After that we explain our proactive approach including tool support by defining a mapping between a lightweight quality model and metrics (section 4). This mostly deals with syntactic aspects, so we introduced simple reviews (section 5) for the mostly semantic and pragmatic aspects.

## 2 Related Work

Model evolution, as we will present, can be discussed closely related to model libraries and model quality. In the following, we present the current understanding of model evolution, model repositories, and model quality.

Model evolution is often investigated as a goal to be achieve automatically by tool support; as it is for software. There are several tools and research prototypes available. First, COPE supports evolution and co-evolution by monitoring changes in an operation based way. These can be applied as editing traces to other models, i.e., forwarded [Herrmannsdoerfer and Ratiu, 2010]. Our approach differs in a regard that we do not trace changes but focus on edit-time changes and their impact on quality aspects. Second, MoDisco [Eclipse, 2012] (hosted with AM3 [Allilaire et al., 2006]) tries to provide means to support evolution of legacy systems applying model-driven ideas. That means, MoDisco is a tool for re-engineering legacy software by means of models and starting a model driven development from gleaned models. Moreover, co-evolution is discussed. We keep to plain model evolution, but the main distinction to our approach is that we want evolution to be guided and directed instead of being aimlessly.

Regarding model repositories one needs to bare in mind their functionality. Often, they allow for querying, conflict resolution, and version management but no more. This means, evolution and co-evolution are not considered. Examples are, first, MOOGLE, a user friendly model search engine offering enhanced querying [Lucredio et al., 2010]. Second, ReMoDD which focuses on comunity building by offering models in a documentary sense to the community [France et al., 2007]. Mind that all of these model libraries do not consider model evolution. Consequently, we have implemented an enhanced model library [Ganser and Lichter, 2013], offering model evolution as presented below.

This evolution support was enhanced by ideas regarding quality in modeling by Moody [Moody, 2005], because we wanted to establish a common understanding of model quality in our library avoiding that "quality is seen as a subjective and rather social than a formally definable property lacking a common understanding and standards" [Moody, 2005]. This is why we took the quality dimensions by Lindland et al. [Lindland et al., 1994] and applied them in our environment. They comprise *syntactic*, *semantic*, and *pragmatic quality* bearing in mind that these quality dimensions can influence each other as presented by Bansiya et al. [Bansiya and Davis, 2002]. Looking at UML models, there exist manifold model qualities. We chose the work of Lange et al [Lange, 2006] to be most suitable and linked it with metrics. There we employed the work of Genero et al. [Genero et al., 2003], Wedemeijer et al. [Wedemeijer, 2001], and Mohagheghi et al. [Mohagheghi and Dehlen, 2009]. Furthermore, we needed means to assess some semantical and pragmatical aspects. Here we root our ideas on reviews but found Fagans approach to heavyweight [Fagan, 1976]. So we subdivided review tasks using an adoption of the thinking hats as proposed by De Bono [De Bono, 2010].

## 3  Quality Staged Evolution

General model evolution is to be distinguished from model evolution in model libraries [Roth et al., 2013] since the purpose differs. While general model evolution is considered aimless, evolution in model libraries does not make sense if it is undirected. This is due to the reuse focus of model libraries and the fact that these models mostly represent a starting point for modeling. Consequently, models in model libraries do not strive for perfection in every possible deployment scenario but rather an eighty percent solution that will need adaption. Still, evolution in model library needs a sound foundation and tool support.

We developed an approach for quality assured model evolution in a gated and staged manner [Roth et al., 2013]. It defines our approach to model evolution in model libraries and how it can be guided. In the remainder of this section, we will shortly describe the quality staged model evolution approach.

### 3.1 Quality and Evolution in Model Libraries

During modeling some parts of a model might seem so generic or generally reusable that a modeler decides to put them in a model library. This means, some parts of the model are extracted, prepared for reuse, and stored in a model library. At the same time, the modeler annotates the extracted model with a simplified specification, we call *model purpose*. It is supposed to grasp the main intention of the model and reflect the general idea in a few words so this model is explained in a complementary way. As all of this is done, we consider this the moment model evolution of this particular model starts.

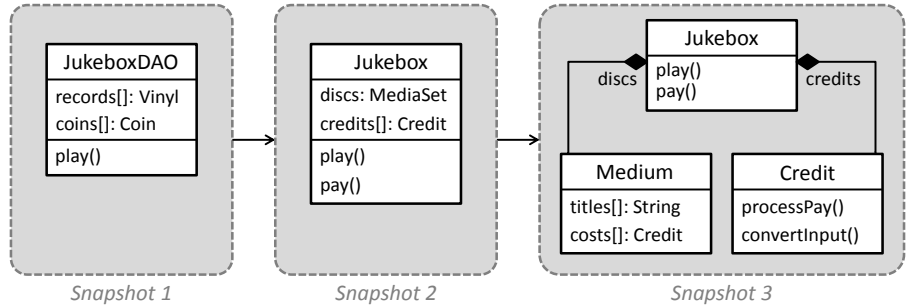

**Fig. 1.** Model Evolution Example

Reasons for model evolution in model libraries can be best explained considering figure 1 as an example. Certainly, the first snapshot of this model is reusable, but some modeling decisions might be questionable. Furthermore, the model is not free from technological details. The "DAO" suffix in the class name is a clumsy leftover that should be removed quickly. Due to that, we offer editing support so models can be overwritten in a versioning style and call a version of a model in our approach a model snapshot.

A few snapshots might be necessary to get a well designed and reusable model. Since all of them are persisted, one can order these snapshots as shown in figure 1 and assign numbers to each snapshot forming an *evolution sequence*.

This evolution sequence can be subdivided into subsequences annotated with stages that make a statement regarding reusability. We conducted a field study about the number and the names and found that "vague", "decent", and "fine" are the best representatives and assigned the colors "red", "yellow", and "green" respectively (cf. figure 2(b)). This is meant to provide an intuitive representation of the model's reusability and the underlying formalities [Roth et al., 2013], since we do not want to bother modelers with the state machine formalizing the states.

The modeler just needs to know the semantics behind each stage. First, a "vague" model might contain some awkward design or leftovers from the originating environment saying: "Be careful reusing this!". Second, a "decent" model is considered reusable in general, but might contain some pragmatic or semantic mismatch between the given purpose and the actual model. This stage is best characterized by: "The devil is in the details." Finally, a "fine" model should be easily reusable and might offer additional information, e.g. template information. So, one could informally characterize it by: "Go ahead and enjoy.".

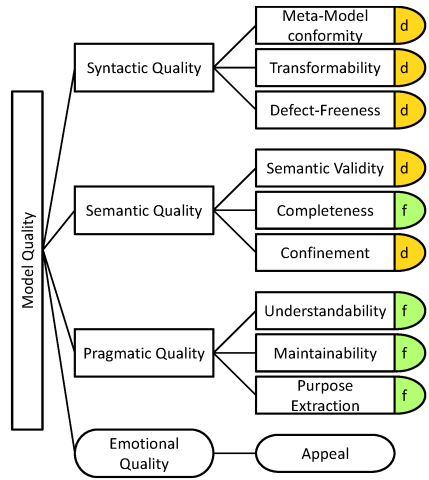
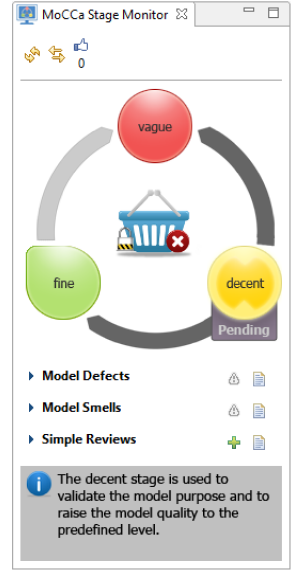

(a) Quality Model and Stages  (b) Eclipse Prototype

**Fig. 2.** Quality Model, Stages, and Prototype similar to [Roth et al., 2013]

The more formal idea behind the quality stages is a quality model that defines criteria (cf. figure 2(a)), which need to be met to gain a certain stage. This is why we talk about quality gates that need to be passed between the stages. In more detail, each quality attribute in figure 2(a) is mapped to a stage in figure 2(b) indicating which quality attributes are required for a certain stage. For example, *completeness* is only required if a model should be regarded "fine", therefore we attached an "f" to that quality attribute in figure 2(a)

Some of the criteria of a quality model might be checked automatically and some might depend on modeler interaction. As a result, the formalization underneath is non-deterministic [Roth et al., 2013], partly because some semantic and most of the pragmatic quality attributes are a matter of subjectivity. For example, contradicting attributes are very unlikely found by tools. If one of the required quality attributes of a gate is not met any more the model loses its status automatically and falls back to the next lower stage.

### 3.2 Quality Measurement Instruments

Evaluating quality attributes of models shows that some of them are automatically assessable and some are not. Clearly, syntactic errors can be found easily by parsers, but completeness is in the eye of the beholder. Consequently, we make a distinction regarding model quality measurement instruments in three categories: *strong*, *medium*, and *weak characteristics*. This classification enables a mapping from model qualities (cf. figure 2(a)) to quality measurement instruments, where each attribute is used to derive a feedback with respect to the attribute name.

Strong characteristics form the strictest type. They can be measured precisely using model metrics. A model metric is formulated with respect to models and provides clear feedback including the reason for the improvement and the suggested solution. For example, a model including a class without a name. Besides, model metrics strong characteristics can be measured with external tools, e.g. EMF validator and EMF generator [Steinberg et al., 2009]. With respect to our quality model in figure 2(a), strong characteristics can be used to derive feedback of the following model quality characteristics: defect-freeness, meta-model conformity, and transformability.

Medium characteristics are based on Fowler's idea of smells [Fowler, 1999]. A smell is something that does not seem to be right and can be measured in some way. For example, a model with hundreds of classes is harder to understand than one with only a few. Such characteristics can be measured with metrics, which define a clear threshold. However, this threshold can be overridden, if the modeler does not agree. Medium characteristics can be used to derive information for confinement, understandability, and maintainability.

Finally, weak characteristics can be compared to hunches. A hunch is something that does not seem to be right because of gut feelings, experience, or intuition. Clearly, it is hard to measure such weak characteristics using metrics. We present simplified reviews in section 5 that enable assessing weak characteristics in a quick and precise way. Such model reviews allow to derive qualitative feedback on semantic validity, completeness, purpose extraction, and appeal.

## 4 Proactive Quality Assessment

Quality measurement instruments and a quality model are used to assess the quality of a model. Such an assessment is, generally, triggered manually at a certain point in time. At this point, the model is analyzed and a report is created, which identifies improvements of the model. Clearly, such improvements can be very vague making the cause for an improvement or a suggested solution hard to understand. Additionally, such events that trigger model assessment are mostly of manual nature, i.e., triggered by someone. Consequently, to prevent long assessment reports with dozens improvement suggestions, such assessment events should be triggered automatically and more importantly periodically.

*Proactive quality guidance* is an approach that triggers assessment events automatically and regards the iterative nature of model creation, i.e., the final model is created in multiple iterations. During model creation the assessment is triggered whenever the model has been changed. The model is analyzed and feedback is presented to the modeler. Because the assessment is triggered when a model is changed, the resulting assessment iterations are kept small avoiding large reports and improvement suggestions. Due to constant and precise feedback during model creation the model evolution is guided and such detected violations with respect to the quality model in section 3.1.

The main parts of proactive quality guidance are (a) automatic and constant assessment of the model, which is currently created, and (b) clear instructions on where the improvements have to be made and why. Automatic assessment is executed when the model is changed but clear and instructive feedback is challenging, because it identifies areas of improvement and their cause and must always be correct. Otherwise, modelers will be annoyed by false feedback. However, the subjective nature of model quality makes it hard to always derive correct feedback without manual interaction.

As the underlying source of information for feedback are quality measurement instruments, we applied the classification of quality measurement instruments, as presented in section 3.2, to structure feedback and, thereby, to loosen up the restriction of always correct feedback. Always correct feedback relies on strong characteristics, which can be measured precisely by using metrics, e.g. if a class has duplicate methods. Furthermore, feedback relies on medium characteristics, which are less precise than strong characteristics, are only suggestions and can be ignored by the modeler. For instance, methods with long parameter lists should be avoided to not pass everything as a parameter. A list of all strong and medium characteristics metrics is listed in [Roth, 2013]. Finally, weak characteristics regard the subjective nature of model quality and, consequently, are hard to measure. In consequence, we present an approach to simplify reviews and to enable measurement of weak characteristics. Such weak quality measurement instruments can then be used to provide feedback.

## 5   Simplified Reviews

Metrics enable proactive quality assessment for strong and medium characteristics but for some weak characteristics proactive quality assessment is difficult. At best heuristics can support modelers but they are unlikely to overrule experience or gut feeling. For example, purpose extraction can be checked partially by keyword comparison but the modeler must have the last word. This is why we researched on simplifying reviews as a means to quickly quality check these and weaker characteristics.

Our result is an approach, we call simplified reviews, that separates different aspects of reviews by altering a technique used in parallel thinking [De Bono, 2010]. These "Six Thinking Hats" provide a separation of concerns for each role which is behind each hat directing tasks clearly. In our simplified reviews this leads to reviews that take no longer than absolutely necessary:

In total five review roles remained because a role for controlling is not necessary for this approach. The hats are designed as follows: A *Yellow Hat Review* (Good points judgment) considers positive aspects of a model and a high number indicates better quality. For example, a review might emphasize that a model is of high benefit in maintenance. The *Black Hat Review* (Bad points judgment) can be regarded as the most known type of reviews. It is used to criticize pointing out difficulties, dangers, defects, or bad design. A black hat review indicates that the corresponding model needs to be patched immediately. A *White Hat Review* (Information) is used to provide information or ask for information, which cannot be gleaned from the model. For example, if a modeler has expertise on limitations of the model, this should be documented by a white hat review. The *Green Hat Reviews* (Creativity) are a means to provide information about possible improvements or new ideas. For a model library this review type is an integral part to foster evolution and to keep modelers satisfied in the long run. Finally, in a *Red Hat Review* (Emotions) a reviewer can express a general attitude in terms of like and dislike. For example, this can be a dislike based on experience that might help improving a model in future.

All of the simple reviews need no more than very simple tool support. A look at figure 2(b) shows an entry that reads "Simple Reviews" with a plus button right next to it. Clicking this button opens a small window that allows to select the type of the simple review and entering additional text. Moreover, the tree editor can be unfolded if there are simple reviews related to this model. Then every review can be inspected and checked "done" or "reopened". This is a bit similar to a very lightweight issue tracking system.

## 6  Proactive Quality Guidance in a Nutshell

General model evolution is to be distinguished from model evolution in model libraries as we briefly discussed above. This is due to unguided evolution being self-defeating for model libraries. Due to that guidance is required to keep models reusable. Moreover, a model library with a focus on reuse puts constraints on a quality model for models that makes it manageable.

All in all, we have shown how models should evolve in model libraries with proactive quality guidance. Therefore, we illustrated how a quality model for model library can be used to guide and stage model evolution in model libraries. To achieve this, we broke down the stages "vague", "decent", and "fine" to quality characteristics which are assured in different ways. While strong characteristics are checked automatically, medium and weak characteristics require user interaction. But this interaction is supported in two ways. On the one hand, for medium characteristics some metrics provide assessments that only need to be judged by a user because certain constraints like thresholds might not hold true for a particular case. On the other hand, for weak characteristics, we introduced simple reviews that allow quick and guided evaluations.

Since all this takes place during editing time, we call this approach proactive quality guidance. But there is more, because a lot of metrics allow to derive recommendations how to fix certain issues. We use other published experience to do so and implemented a prototype that realizes the entire approach. It looks simple and clean, because we tried to avoid as much noise for modelers as possible so the modeler does not get distracted while modeling.

# Acknowledgements

We would like to thank all our anonymous reviewers for their comments and effort. We would also like to thank all our participants in our surveys and studies.

# References

[Allilaire et al., 2006] Allilaire, F., Bezivin, J., Bruneliere, H., and Jouault, F. (2006). Global Model Management in Eclipse GMT/AM3. In *Proceedings of the Eclipse Technology eXchange workshop (eTX) at the ECOOP 2006 Conference*.

[Bansiya and Davis, 2002] Bansiya, J. and Davis, C. (2002). A hierarchical model for object-oriented design quality assessment. *IEEE Transactions on SE*.

[Chaudron et al., 2012] Chaudron, M., Heijstek, W., and Nugroho, A. (2012). How effective is UML modeling? *Software and Systems Modeling*, pages 1–10.

[De Bono, 2010] De Bono, E. (2010). *Six Thinking Hats*. Penguin Group.

[Eclipse, 2012] Eclipse (2012). MoDisco. http://www.eclipse.org/MoDisco/.

[Fagan, 1976] Fagan, M. E. (1976). Design and code inspections to reduce errors in program development. *IBM Systems Journal*, 15(3):182–211.

[Fowler, 1999] Fowler, M. (1999). *Refactoring: Improving the Design of Existing Code (Object Technology Series)*. Addison-Wesley Longman, Amsterdam.

[France et al., 2007] France, R., Bieman, J., and Cheng, B. (2007). Repository for Model Driven Development (ReMoDD). In Kuehne, T., editor, *Models in Software Engineering*, volume 4364 of *LNCS*, pages 311–317. Springer Berlin / Heidelberg.

[Ganser and Lichter, 2013] Ganser, A. and Lichter, H. (2013). Engineering Model Recommender Foundations. In *Modelsward 2013, Proceedings of the 1st International Conference on Model-Driven Engineering and Software Development, Barcelona, Spain,19.-21- February 2013*, pages 135–142. SCITEPRESS.

[Genero et al., 2003] Genero, M., Piattini, M., Manso, M. E., and Cantone, G. (2003). Building uml class diagram maintainability prediction models based on early metrics. In *IEEE METRICS*, pages 263–. IEEE Computer Society.

[Herrmannsdoerfer and Ratiu, 2010] Herrmannsdoerfer, M. and Ratiu, D. (2010). Limitations of automating model migration in response to metamodel adaptation. In Ghosh, S., editor, *Models in Software Engineering*, volume 6002 of *Lecture Notes in Computer Science*, pages 205–219. Springer Berlin Heidelberg.

[Lange and Chaudron, 2005] Lange, C. F. and Chaudron, M. R. (2005). Managing model quality in uml-based software development. *Software Technology and Engineering Practice, International Workshop on*, 0:7–16.

[Lange, 2006] Lange, C. F. J. (2006). Improving the quality of uml models in practice. In *Proceedings of the 28th international conference on Software engineering*, ICSE '06, pages 993–996, New York, NY, USA. ACM.

[Lindland et al., 1994] Lindland, O., Sindre, G., and Solvberg, A. (1994). Understanding quality in conceptual modeling. *Software, IEEE*, 11(2):42 –49.

[Lucredio et al., 2010] Lucredio, D., de M. Fortes, R., and Whittle, J. (2010). MOOGLE: a metamodel-based model search engine. *Software and Systems Modeling*, 11:183–208.

[Melnik, 2004] Melnik, S. (2004). *Generic Model Management: Concepts and Algorithms*. Lecture Notes in Computer Science. Springer.

[Mens et al., 1999] Mens, T., Lucas, C., and Steyaert, P. (1999). Supporting disciplined reuse and evolution of UML models. In Bezivin, J. and Muller, P.-A., editors, *Proc. UML'98 - Beyond The Notation*, volume 1618 of *Lecture Notes in Computer Science*, pages 378–392. Springer-Verlag. Mulhouse, France.

[Mohagheghi and Dehlen, 2009] Mohagheghi, P. and Dehlen, V. (2009). Existing model metrics and relations to model quality. In *Proceedings of the Seventh ICSE conference on Software quality*, WOSQ'09, pages 39–45, Washington, DC, USA. IEEE.

[Moody, 2005] Moody, D. L. (2005). Theoretical and practical issues in evaluating the quality of conceptual models: current state and future directions. *Data Knowl. Eng.*, 55(3):243–276.

[Roth, 2013] Roth, A. (2013). A Metrics Mapping and Sources. http://goo.gl/ruqFpi.

[Roth et al., 2013] Roth, A., Ganser, A., Lichter, H., and Rumpe, B. (2013). Staged evolution with quality gates for model libraries. In *1st International Workshop on:(Document) Changes: modeling, detection, storage and visualization, September 10th, 2013, Florence, Italy*.

[Stachowiak, 1973] Stachowiak, H. (1973). *Allgemeine Modelltheorie*. Springer-Verlag.

[Steinberg et al., 2009] Steinberg, D., Budinsky, F., Paternostro, M., and Merks, E. (2009). *EMF: Eclipse Modeling Framework 2.0*. Addison-Wesley, 2nd edition.

[Wedemeijer, 2001] Wedemeijer, L. (2001). Defining metrics for conceptual schema evolution. In *Selected papers from the 9th International Workshop on Foundations of Models and Languages for Data and Objects, Database Schema Evolution and Meta-Modeling*, FoMLaDO/DEMM 2000, pages 220–244, London, UK. Springer.